\documentclass[12pt]{iopart}

\usepackage{graphicx}
\usepackage{tabularx}
\usepackage[square,numbers,sort&compress]{natbib}
\usepackage{setstack}
\usepackage{iopams}
\usepackage{amsfonts}

\newcommand{\textcolor}[2]{#2}

\usepackage{geometry} 

\def\newblock{\hskip .11em plus .33em minus .07em}

\newcommand{\appropto}{\mathrel{\vcenter{
  \offinterlineskip\halign{\hfil$##$\cr
    \propto\cr\noalign{\kern2pt}\sim\cr\noalign{\kern-2pt}}}}}

\renewcommand{\vec}[1]{\bi{#1}} 

\begin{document}

\title[Tangled vortex lines in 3D random wave fields]{Geometry and
  scaling of tangled vortex lines in three-dimensional random wave
  fields}
\author{A J Taylor and M R Dennis}
\address{H H Wills Physics Laboratory, University of Bristol, 
    Tyndall Avenue, Bristol, BS8 1TL, UK}
\eads{\mailto{alexander.taylor@bristol.ac.uk}, \mailto{mark.dennis@bristol.ac.uk}}

\begin{abstract}
  The short- and long-scale behaviour of tangled vortices (nodal
  lines) in random three-dimensional wave fields is studied via
  computer experiment.  The zero lines are tracked in numerical
  simulations of periodic superpositions of three-dimensional complex
  plane waves.  The probability distribution of local geometric
  quantities such as curvature and torsion are compared to previous
  analytical and new Monte Carlo results from the isotropic Gaussian
  random wave model.  We further examine the scaling and
  self-similarity of tangled wave vortex lines individually and in the
  bulk, drawing comparisons with other physical systems of tangled
  filaments.
\end{abstract}

\pacs{02.40.Hw,03.65.Vf,05.40.-a}

\section{Introduction}
\label{sec:introduction}

Tangles of filaments filling 3-dimensional space occur generically in
physics.  In the theory of complex-valued scalar waves, such tangles
occur as the \emph{wave vortices} (also called nodal lines, phase
vortices or wave dislocations) \cite{dennis2009,nye1974} in linear
superpositions of complex waves whose directions and phases are
uniformly random but whose wavelength $\lambda$ is fixed.  Informally,
these tangles of nodal filaments exist in any monochromatic component
of 3-dimensional (3D) noise: in the light~\cite{freund2000} or sound field in the space
around us, or in an evolving and interfering matter wave without any
overall direction of propagation.  As such, these isotropic random
wave fields are the most natural model of wave chaos \cite{berry1977},
and lead to tangles of random vortices in models of chaotic 3D cavity
modes~\cite{berggren2009}. These tangles are similar to initial conditions for
models of cosmic strings~\cite{halperin1981}, and isotropic, fully
developed optical speckle patterns~\cite{oholleran2008}.  As they are
nodal lines, no energy is associated with wave vortices, so their
geometry and physics is distinct from other well-studied filamentary
physical systems such as elastic rods, superfluid vortices, or defects
in ordered media, despite the fact that all these systems form
morphologically similar tangles.  In this paper we investigate the
small- and large-scale geometric structure of tangled wave vortices
from computer simulations of random wave fields.

Although the geometric structure of tangled nodal lines in random wave
superpositions is hard to characterise rigorously, several previous
attempts have been made to approach them using analytic and numerical
methods.  On small lengthscales ($\lesssim \lambda$) the amplitude of
a chaotic random wavefield is a smooth complex function of position,
so the nodal lines, as complex zero level sets, are themselves smooth
space curves.  Statistical quantities such as the density of vortex
lines per unit volume~\cite{halperin1981,berry2000,ishio2001} and the
probability distribution of the vortex lines' curvature
\cite{berry2000} can be computed analytically, treating the wave field
as an isotropic Gaussian random function.  Calculation of the
statistical distribution of other geometric quantities, such as the
lines' torsion, seems analytically intractable by these
methods~\cite{mrdthesis}.

At larger lengthscales, properties of interest cannot be found
analytically from differential geometry and the Gaussian random model,
and computer experiments are necessary.  Computer simulations of the
nodal structure of random optical fields inside periodic 3D cells have
indicated that, at distances much larger than the coherence length
(comparable to $\lambda$), a typical random nodal line looks like a
Brownian random walk~\cite{oholleran2008}.

\begin{figure}
  \includegraphics{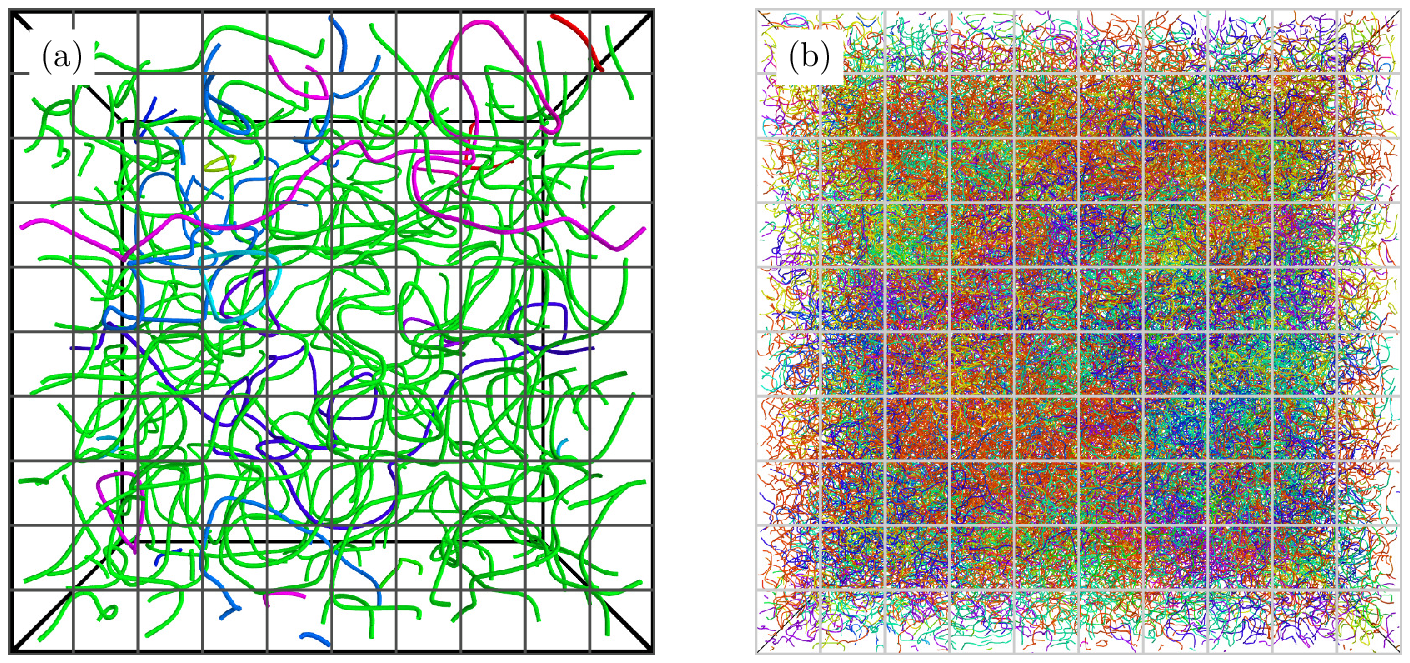}
  \caption{\label{fig:3dcell} Examples of vortex tangles of random
    wave superpositions in periodic cells.  Cells have side length (a)
    $4.33\lambda$ and (b) $21.7\lambda.$ Each unique vortex line is
    coloured differently.  The grid lines are spaced at $25\times$ the
    basic numerical sampling resolution (grid spacings of (a)
    $0.43\lambda$ and (b) $2.17\lambda$).  As described in Section
    \ref{sec:numerics}, the real numerical sampling resolution of (a)
    is at least $175$ points per wavelength.  }
\end{figure}

Our aim here is to explore, using large-scale computer simulations of
random solutions of the Helmholtz equation in large periodic 3D cells,
the crossover in the geometric behaviour of the tangle of vortices
between the local regime where the vortices are smooth space curves,
and longer lengths where the lines and tangles display various
measures of statistical self-similarity.  Our simulations employ
superpositions of random waves with periodic boundary conditions,
where the vortex tangle is periodic with a cubic fundamental cell
(Figure \ref{fig:3dcell}), which are good approximations of wavefields
whose statistics are genuinely statistically invariant to rotation.

More quantitatively, the wave fields we study are drawn from the
ensemble of superpositions of complex-valued plane waves with random
amplitudes and directions, i.e.~
\begin{equation}
  \psi(\bi{r}) = \sum_{\bi{k}} a_{\bi{k}} \exp(\rmi \bi{k}\cdot\bi{r})~,
  \label{eq:rwm}
\end{equation} 
where $\bi{r} = (x,y,z),$ the wavevectors $\vec{k}$ have the same
fixed wavenumber $|\vec{k}| \equiv k = 2\pi/\lambda$ so $\psi$ in
(\ref{eq:rwm}) satisfies the 3D Helmholtz equation $\nabla^2 \psi +
k^2 \psi = 0$, and the complex amplitudes $a_{\vec{k}}$ are
independent, identically distributed circular Gaussian random
variables.  In our simulations, the $\vec{k}$ are chosen to ensure
cubic periodicity of the nodal pattern in real space; we impose that
each $\vec{k} = 2\pi \vec{c} / N$ for a fixed integer $N$, with
$\vec{c}$ a 3-vector of odd integers such that
$|\vec{c}|=\sqrt{3}N$.  Thus, $\psi(\vec{r}) = \pm\psi(\vec{r}+
\vec{d}N/2)$ for any $\vec{d}$ in $\mathbb{Z}^3,$ implying the zero
pattern is periodic over half the period of the wavefield in each of
the three periodic direction of the cell (i.e.~one octant of a field
cell).  It is this periodicity of a cubic nodal line cell to which we
refer through throughout the text.

Varying $N$ whilst keeping $\lambda$ fixed may be considered as
changing the size of the periodic cell at fixed energy, making the
vortex structure in cells of different sizes directly comparable.
Equation (\ref{eq:rwm}) denotes a mathematically natural physical
system, and in the limit $N \to \infty$ is determined by only one
physical length parameter---the wavelength $\lambda$---and so the
geometric properties of the nodal lines of $\psi(\vec{r})$ are purely
inherent features of isotropic random wave interference.

Our simulations employ three distinct cell sizes, with side lengths
$4.3\lambda$, $7.8\lambda$ and $21.7\lambda$ (corresponding to
$N=5, 9,$ and $25$ respectively).  Samples of the smallest and largest
vortex cells are shown in Figure \ref{fig:3dcell}.  In the
mathematically ideal case, the cell would be infinitely large for
fixed $\lambda,$ and in the limit the sum would be over an isotropic,
continuous distribution of wavevectors, captured analytically by the
Gaussian random wave model (e.g.~\cite{berry2000}).  In practice the
periodicity condition limits the sum to a smaller number of component
wavevectors (effectively between $50$ to $100$; values of $N$ are
chosen to maximise these), which reduces statistical isotropy and may
constrain some geometric quantities.  A way this may be gauged is via
the vortex densities in two and three dimensions, compared with their
counterparts calculated analytically from the isotropic random wave
model \cite{berry2000}: the ideal isotropic wave vortex tangle penetrates an
arbitrary 2D plane at points with a density $2\pi/3\lambda^2 \approx
2.09/\lambda^2,$ whilst in 3D the density of vortex \emph{arclength}
per unit volume is twice this, $4\pi/3\lambda^2 \approx
4.19/\lambda^2.$
In our simulations with $N=25$, the 2D density is
$(2.12\pm0.03)/\lambda^2$ in planes perpendicular to the axes of cubic
periodicity, and $(2.14\pm0.03)/\lambda^2$ through planes passing
through the diagonal of a periodic cell face -- both consistent with
the theoretical result, and isotropic within the error bounds.  The
3D vortex densities from simulation are $(4.45\pm0.03)/\lambda^2$,
$(4.61\pm0.04)/\lambda^2$ and $(4.49\pm0.03)/\lambda^2$ in cells with
$N=5, 9, 25$ respectively, now not quite fully consistent with the
theoretical result.  We suspect this reflects the periodicity
constraint, which forces vortex lines into loops more often than in
the fully isotropic model (discussed further in
Section~\ref{sec:scaling}), but does not appear significantly to
affect our main results.

The properties of wave vortex tangles in this simple system will be
compared with tangled filaments in other physical systems such as
vortex filaments in turbulent superfluids, and polymer melts.  They
are also of interest by comparison with the high energy complex
eigenfunctions of chaotic 3D cavities, for which functions of the form
(\ref{eq:rwm}) are conjectured to be a good model, and for which the
vortices in non-time-reversal-symmetric
eigenfunctions~\cite{berry1986} and resonances~\cite{hohmann2009} have
been studied in 2D. The wave vortices are the 3D nodal counterpart to
the nodal lines bounding nodal domains in the 2D real chaotic case,
whose behaviour has been extensively
characterised~\cite{berry1977,blum2002,bogomolny2002,monastra2003,bogomolny2007}.
In particular, nodal line behaviour in the isotropic random wave model
has been compared with numerically computed high-energy eigenfunctions
in billiards. The results in this paper can be considered as some
basic characterization of these nodal lines in 3 dimensions, including
preliminary comparison between the isotropic model and complex
eigenfunctions in the 3-torus (also see~\cite{gnutzmann2014}).
Physically, the vortex lines are important in wave chaos, for instance
in dominating the flow of the current~\cite{berggren2009}.

The structure of this paper is as follows.  In the following Section,
details are given of our numerical implementation of the random wave
cells and the tracking of vortex lines in them, especially the new
supersampling technique we employ.  In Section~\ref{sec:geometry}, we
apply the numerical tracking on the short scale to extract the
behaviour of curvature, torsion and their rate of change along a
single vortex line, comparing with previous analytic results of
curvature and new Monte Carlo (MC) calculations for torsion.  Section
\ref{sec:scaling} follows with an investigation of structure and
scaling at larger scales, including the fractal behaviour of
individual lines, the scaling of size and number of closed loops, and
the fractal dimension of the tangle itself.  We conclude with brief
discussion.

\section{Numerical random wave superpositions}
\label{sec:numerics}

In this Section we describe how the numerical tangle of vortex lines
is extracted from the periodic random wave superpositions described
above.  The method follows previous numerical experiments
\cite{oholleran2008}, by sampling the field at points on a 3D cubic
grid within the periodic cell, and searching for $2\pi$ net phase
changes around the four sides of each square plaquette face in the
cubic lattice.  (This is reminiscent of the so-called $\mathbb{Z}_3$
lattice model in which the phases are themselves
discrete~\cite{vachaspati1984}.)  Within each cube bounded by six such
plaquettes, there may be a vortex denoted by a $2\pi$ phase change
oriented {\em into} the cube, and another vortex with a $2\pi$ phase
change oriented {\em out} of the cube.  In this way the tangle of
vortex lines can be built up through the volume.  Readers uninterested
in further details of vortex tracking may skip to \ref{sec:geometry}.

\begin{figure}
  \includegraphics{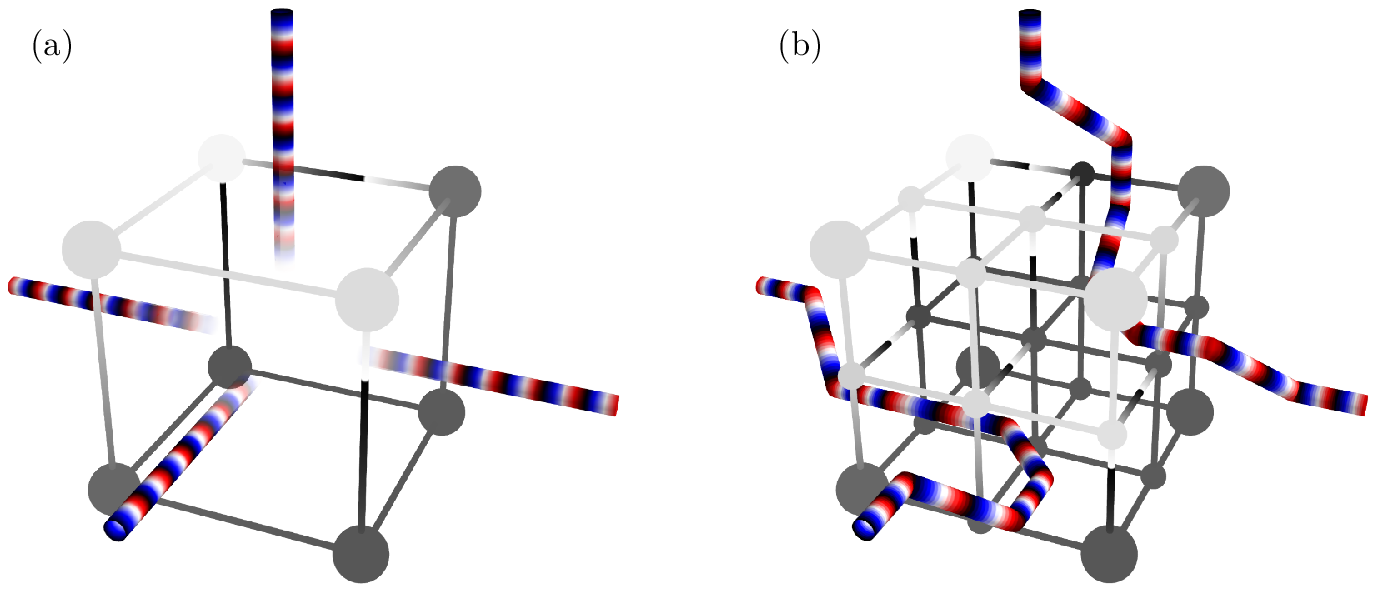}
  \caption{\label{fig:resampling} Demonstration of the supersampling
    vortex tracking algorithm.  Each sphere at a vertex of the
    sampling grid represents a phase voxel, with phase indicated by
    greyscale.  Lines between the voxels are coloured following a
    linear interpolation of phase between them (following the shortest
    change mod $2\pi$).  The patterned lines represent vortex lines,
    the increase red-white-blue denoting the direction of the vortex
    line, from a right-handed increase of phase.  These lines enter or
    leave the cell through faces around which the total phase change
    is $2\pi$.}
\end{figure}

The new feature of the tracking algorithm implemented here is that the
basic sampling lattice is \emph{resampled} in regions where the
spatial details of the tangle are smaller than the resolution of the
sampling grid.  This is important because the basic algorithm cannot
resolve vortex topology when two or more vortex lines enter the same
sampling grid cell, shown in Figure \ref{fig:resampling}(a).  In
addition, the local vortex line geometry is poorly recovered unless
the sampling grid resolution is high on the scale of the curvature of
the line, but the computation time and memory requirements scale with
the cube of the grid size.  This ambiguity in resolving by vortex
topology was approached in previous studies by different techniques;
in the optical wave model of~\cite{oholleran2008} using physical
arguments, and in the discrete model of~\cite{vachaspati1984} by an extra
random variable determining the topological assignment, or moving to a
different 3D grid which does not permit ambiguity
\cite{hindmarsh1995}. 

We believe our recursive resampling algorithm is more robust given the
physics of our random wave model; whenever the local geometry or
topology within a sampling cell cannot be resolved, the local phase
structure in that cell is sampled again at a progressively higher spatial
resolution until all local vortices can be fully distinguished,
enforcing that no two vortex lines intersect the same cubic grid cell
in the new resampled lattice.

Figure \ref{fig:resampling} shows an example of this process; the
phase is initially sampled at the vertices of Figure
\ref{fig:resampling}(a) revealing only that two vortices enter and
leave the cell through the four cell sides exhibiting a $2\pi$ phase
change. This is not enough information to
distinguish the sub-cell topology of the vortices.  After
resampling, the paths of both vortex lines in Figure
\ref{fig:resampling}(b) are tracked without ambiguity, as they do not
approach closely on the scale of the resampled grid.  This fully
solves the problems of both topological resolution and numerical
efficiency by focusing computational effort on areas where vortices
cannot be resolved properly, sampling only fairly sparsely elsewhere
(say in volumes where there are no vortices).

This method resolves vortex lines very precisely on the sub-wavelength
scale.  However, the individual linear segments do not accurately
reproduce the conformation of the vortex filament, as it is assumed
that each vortex line segment passes through the centre of the face
where it are detected.  The local fit is improved via three further
numerical steps.  First, once the vortex lines have been located, the
resampling procedure is applied again to all the cells through which
the vortex line passes, even if the topology is already well
distinguished.  This accurately traces the shape of the vortex curves
without expending further numerical resources on the voids between
longer vortices. In principle, extremely small vortex loops below the
sampling cell may not be detected, but these are extremely rare and
contribute a negligible amount to the total vortex arclength.

 Secondly, rather than assuming the vortex line penetrates
the centre of the grid face, we instead treat the face plane as a new,
two-dimensional vortex location problem, and the intersection point is
located by resampling the face with a 2D square grid of $15\times$ the
local 3D resampling resolution, and locating the square plaquette of
this new grid around which there is still a $2\pi$ phase change.  The
centres of these new plaquettes in the 2D face are joined by straight
lines in the 3D grid to approximate the real vortex curve.

The result of these procedures is a representation of
every vortex filament as a piecewise-linear space curve.  Any
remaining numerical noise and artefacts of the numerical sampling
lattice are removed by downsampling this representation, keeping only
every $n$th vertex, and fitting with a piecewise polynomial spline
curve interpolation.  In case the rigid sublattice sampling introduces
high frequency noise to the fitting polynomial, the spline curve fit
is required only to closely approximate the original vertices.  We
require that the sum of square distances between original and
interpolated points does not exceed some small value; $\sum_v
(\vec{r}_v - \vec{r}_i) < \delta L$, where $\vec{r}_{\mathrm{v}}$
represents the vertices of the original piecewise linear curve,
$\vec{r}_{\mathrm{i}}$ is the interpolated spline curve at each of
these points, $\delta$ is a fitting parameter, $L$ is total length of
the current vortex curve and the sum is over the vertices.  This
procedure resembles that of~\cite{baggaley2011} in tracing superfluid
vortices, though here we are concerned with accurately recovering the
small scale geometry rather than removing noise on larger scales.

There are many possible choices of downsampling and interpolation
parameters.  We choose to downsample to every third vertex,
and to set the fitting parameter $\delta=0.0001$.
These choices are such that the recovered geometry is stable to other
values of each parameter.

The results of Section \ref{sec:geometry} on the local geometry of
vortex lines are taken from 2628 sample wavefunctions with $N=5$ (such
as Figure \ref{fig:3dcell}(a)), whose cubic periodic cell has volume
$(4.33\lambda)^3$.  On average each cell contains $(361 \pm
3)\lambda$ total vortex length.  The wavefunction is sampled on an
initial grid of $250^3$ points, a real space resolution of
$0.017\lambda$, and we enforce a minimal resampling of $7\times$
along vortices for a real resolution of $0.0025\lambda$ along vortex
curves.  Tracing vortices in a single sample wavefunction takes around
3 hours on a typical desktop machine, and the resulting vortex curves
are sets of piecewise linear segments at a resolution of approximately
$400$ vertices per wavelength.

Section \ref{sec:scaling} investigates fractality over a wide range of
scales, so these results are taken instead with 650 cells with $N=9$
and 47 cells with $N=25$ (e.g. Figure \ref{fig:3dcell}(b)).  These
have periodic domain $(7.8\lambda)^3$ and $(21.7\lambda)^3$
respectively, and are both sampled with an initial real space
resolution of $0.087\lambda$.  The results from these larger cells are
not smoothed and resampled further, as measurements of large scale
fractality are not sensitive to small-scale numerical noise.  This
approach, trading local precision for speed, takes approximately
minutes to analyse each wavefunction if $N=9$, or about $2$ hours to
analyse each wavefunction when $N=25$ (i.e.~$(2180\pm20)\lambda$ or
$(45500 \pm 400)\lambda$ respectively).

\section{Local geometry of single vortex curves}
\label{sec:geometry}

At the scale of a wavelength or less, we approach the geometric
structure of the tangled melange of vortex filaments by considering
the statistical shape of a single vortex line, ignoring the line's
global topology and other lines nearby.  At the scale of $\lambda,$
the shape of the zero lines is limited by the Helmholtz equation, and the
vortex lines are smooth space curves parametrised by
\emph{curvature} $\kappa$ and \emph{torsion} $\tau,$ which, as
functions of arclength $s$ uniquely determine the shape of the curve
\cite{moderndifferentialgeometry}, and are defined (as in
\ref{appendix:fs}) in terms of derivatives along the curve with
respect to $s.$ As natural measures of the tangling of a single space
curve, $\kappa$ and $\tau$ have been studied in other systems
including superfluids \cite{baggaley2011} and even discrete lattice
models \cite{bickis1998}.
We investigate the statistical distribution of these quantities via
the isotropic Gaussian random wave model in which the probability
distribution for curvature is available analytically~\cite{berry2000},
and the distributions of mathematically more complicated quantities
can be obtained through Monte Carlo integration.

The curvature and torsion occur as part of the \emph{Frenet-Serret}
formalism of space curves, outlined in \ref{appendix:fs}.  In our
system, the arclength along wave vortex lines is naturally expressed
in terms of the \emph{vorticity} vector $\boldsymbol{\omega} \equiv
\frac{1}{2}\mathrm{Im}\nabla \psi^* \times \nabla \psi$ (evaluated on
nodal lines), whose direction gives the tangent to the vortex curve at
that point (with respect to a right-handed increase of phase), leading
to the expressions (evaluated on nodal lines)
\begin{eqnarray}
  \kappa &= \frac{|\boldsymbol{\omega} \times (\boldsymbol{\omega} \cdot \vec\nabla) \boldsymbol{\omega}|}{\omega^3}~,
  \label{eq:curvature} \\
  \tau &= \frac{\boldsymbol{\omega} \times (\boldsymbol{\omega}\cdot\vec\nabla) \boldsymbol{\omega} \cdot (\boldsymbol{\omega} \cdot \nabla)^2 \boldsymbol{\omega}}
  {|\boldsymbol{\omega} \times (\boldsymbol{\omega} \cdot \vec\nabla) \boldsymbol{\omega}|^2}~. \label{eq:torsion}
\end{eqnarray}
The curvature therefore depends on second derivatives of the field,
and the torsion on third derivatives.  The numerator of $\kappa$
occurs in the denominator of $\tau,$ suggesting a statistical
anticorrelation discussed further below.

Informally, the curvature and torsion of a space curve may be
understood in terms of a local fit after a suitable translation and
rotation to a \emph{helix} of radius $r$ and pitch $c:$ $\vec{a}(t) =
(r \cos(t), r \sin(t), ct)$ (a helical vortex may be realised in a
wave of the form of a perturbed screw
dislocation~\cite{nye1974,mrdthesis}).  The curvature and torsion of
the helix are the same at all points: $\kappa = r/(r^2 + c^2)$ and
$\tau = c / (r^2+c^2).$ The local radius of curvature is $1/\kappa >
r;$ in the limit $c \to 0,$ $\kappa \to 1/r$ and the curve is locally
a planar circular arc.  In the opposite limit $r \to 0,$ the helix
becomes a straight line and $\tau \to 1/c.$ This limit is therefore
singular (the straight line has an arbitrary value of torsion, which
is technically defined only when $\kappa > 0$).  This suggests that
when, as here, physical considerations do not bound $\tau,$ it can
have very large fluctuations and is numerically ill-behaved when the
curve is locally almost straight.

\begin{figure}
  \centering
  \includegraphics{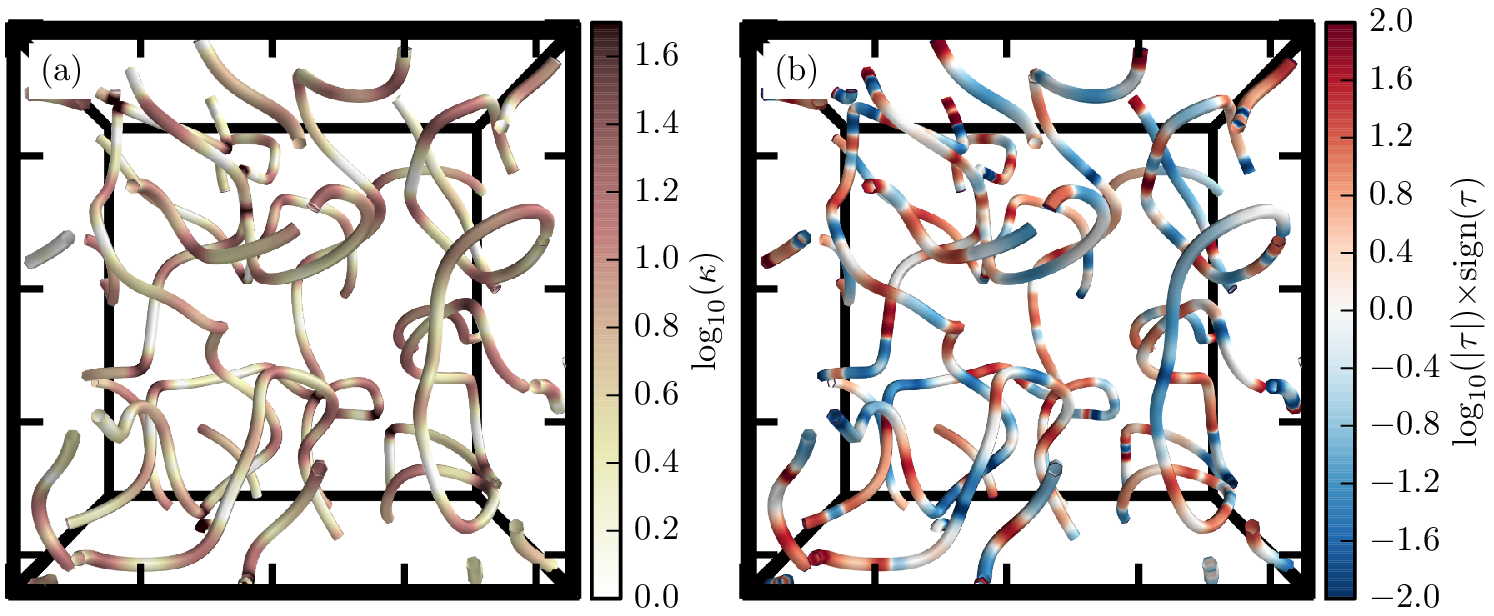}
  \caption{
  \label{fig:fs_boxes} 
  Curvature and torsion of a vortex tangle.  Vortex lines are coloured
  by (a) curvature and (b) torsion, determined numerically.  In each
  case, the same $(2.17\lambda)^3$ volume of one of the cells
  described in Section \ref{sec:numerics} is shown, with the axis
  ticks marking $0.5\lambda$ increments.  }
\end{figure}

Figure \ref{fig:fs_boxes} shows a simulated volume of vortex tangle,
in which the vortex lines are coloured according to $\kappa$ and
$\tau.$ Both quantities exhibit rapid changes on the sub-wavelength
scale.  By eye the curvature is usually small, but with localised
regions of much higher curvature, particularly where pairs of vortex
lines approach closely.  The torsion is also often small with strongly
concentrated regions, but these tend to occur in the less sharply
curved segments of the vortex lines, indicating that the plane of
curvature rotates about the vortex axis when the local curvature is
small.

In other 3D systems with filamentary defect lines, the distributions
of curvature and torsion have direct physical implications, relating
to the local dynamics (as for superfluids~\cite{kivotides2004}) or
energetics (as for defects in liquid crystals~\cite{kamien2002}).
In our case of linear wave superpositions, the wave vortices do not
carry energy and are temporally static, and furthermore do not depend
on the choice of any physical parameter except the wavelength.

\subsection{Curvature}
\label{curvature}

The curvature probability density function (PDF) for vortex lines in
the continuum random wave model was found in \cite{berry2000} to have
the Cauchy-like distribution
\begin{equation}
  P(\kappa) = \frac{3^{5/2}\kappa \kappa_{\mathrm{c}}^3}{(3\kappa_{\mathrm{c}}^2 + \kappa^2)^{5/2}~},
  \label{eq:curvature_pdf}
\end{equation}
where, for fields satisfying the Helmholtz equation, the
characteristic curvature $\kappa_c = 4\pi/\sqrt{45}\lambda \approx
1.87/\lambda.$ Thus the peak of the distribution is at
$2\pi/\sqrt{15}\lambda \approx 1.62/\lambda,$ and its first two
moments are
\begin{eqnarray}
  \langle \kappa \rangle = \sqrt{3}\kappa_c \approx 3.24/\lambda~, \\
  \langle \kappa^2 \rangle = 2\langle\kappa\rangle^2 = 6\kappa_c^2 \approx 21.1 / \lambda^2~,
\end{eqnarray}
whereas higher moments diverge. The values of these moments indicate
that the lengthscales associated with vortex curvature are certainly
sub-wavelength.

\begin{figure}
  \centering
  \includegraphics{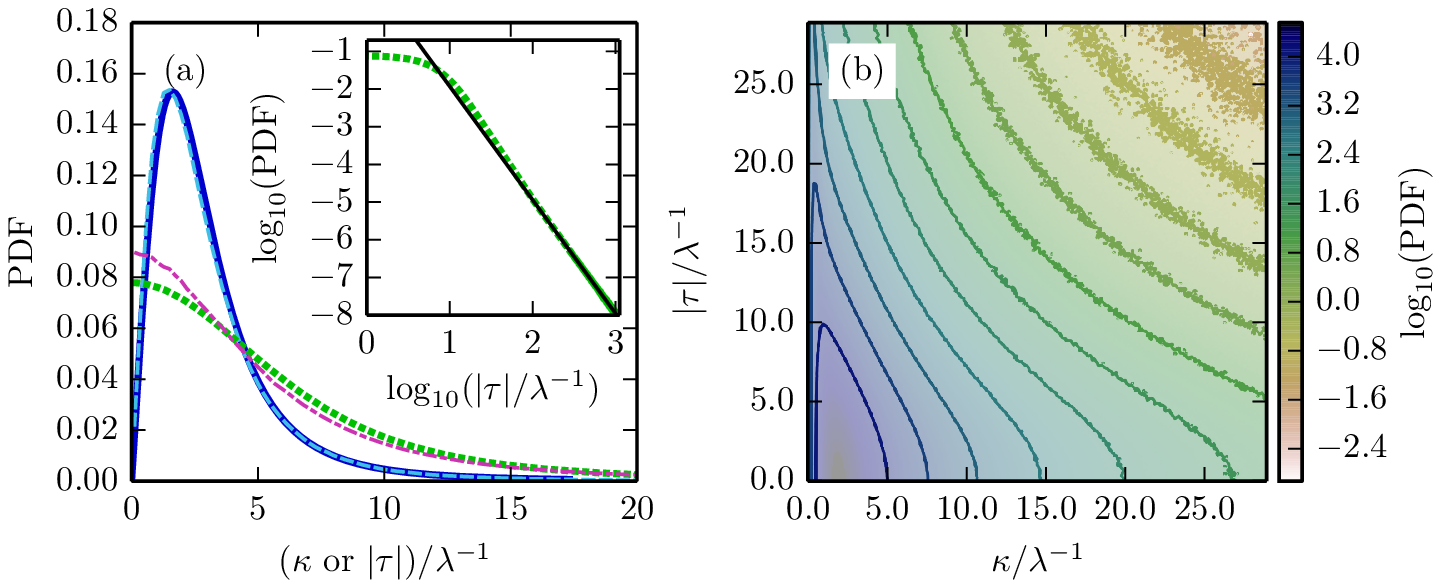}
  \caption{
   \label{fig:fs_pdfs} 
    Curvature and torsion PDFs from the Gaussian random wave model and numerical simulations of random waves.  
    (a) PDFs for curvature and torsion from simulations, analytic results and Monte Carlo integration; (b) joint PDF $P(\kappa,\tau)$ drawn from Monte Carlo integration.  
    In (a), $P(\kappa)$ is determined by (\ref{eq:curvature_pdf}) (\textcolor{paperdarkblue}{\rule[0.4ex]{3mm}{1mm}}), and from the random wave simulations (\textcolor{paperlightblue}{\rule[0.4ex]{1.25mm}{.5mm}\rule[0.4ex]{0.3mm}{0pt}\rule[0.4ex]{1.25mm}{.5mm}}).  
    The torsion PDF is found by Monte Carlo integration (\textcolor{paperlightgreen}{\rule[0.4ex]{0.7mm}{1mm}\rule[0.5ex]{0.3mm}{0pt}\rule[0.4ex]{0.7mm}{1mm}\rule[0.4ex]{0.3mm}{0pt}\rule[0.4ex]{0.7mm}{1mm}}) (described in Section \ref{sec:torsion}), and in the simulations (\textcolor{paperpurple}{\rule[0.4ex]{1.9mm}{0.5mm}\rule[0.4ex]{0.3mm}{0pt}\rule[0.4ex]{0.6mm}{0.5mm}}).  
    Inset: Log-log plot indicating how $\tau$ scales for larger
    values, as discussed in the text of Section \ref{sec:torsion}.
    All Monte Carlo results come from $6\times10^{10}$ points. 
  }
  \end{figure}

The curvature PDF extracted from our numerical simulations is shown in
Figure \ref{fig:fs_pdfs}.  The agreement with (\ref{eq:curvature_pdf})
is very good, albeit with a slight discrepancy at small $\kappa$ where
the vortex tracing is imperfect and detail is lost by the linear
piecewise approximation and smoothing.  The peak of the recovered
distribution is at $\kappa=(2.71 \pm 0.05)/\lambda$, a shift of
$3.4\%$.

  From the random wave simulations, we also recover the expectations
  and moments of the distribution
  \begin{eqnarray}
    \label{eq:curvaturemoment}
    \langle \kappa \rangle = (3.18\pm0.06) / \lambda~, \\
    \label{eq:curvaturesqmoment}
    \langle \kappa^2 \rangle = (18.9\pm0.4)/\lambda^2 = 1.87 \langle\kappa\rangle^2~,\\
    1 / \sqrt{\langle \kappa^2 \rangle} = (0.230\pm0.002) \lambda~.
  \end{eqnarray}
  These are comparable to the analytic results shifted to slightly
  lower curvatures, with the surprising exception that expected
  relationship $\langle \kappa^2 \rangle = 2 \langle \kappa \rangle^2$
  is relatively strongly affected by the shift.

  A fraction of the vortex line length in the simulation occurs as
  small loops (with size of around a wavelength or less), as
  anticipated in \cite{berry2007}.  As these loops have curvature at
  least $2\pi/$radius over their length, they have
  uncharacteristically high curvature for the distribution.  However,
  these loops account for only a small fraction of the overall vortex
  line length in the sample, and the PDF would be reproduced well even
  if only longer vortex lines were sampled.

  \subsection{Torsion}
  \label{sec:torsion}

  The torsion of a nodal line in a random wave field is less easy to
  interpret than its curvature.  As can be seen in Figure
  \ref{fig:fs_boxes}, $\tau$ varies rapidly over lengthscales
  comparable to or shorter than those of $\kappa$ and, as discussed
  above, may be large even where a vortex filament is almost straight
  -- it describes instead the local twisting of the Frenet frame about
  the tangent axis.

  The statistical properties of the torsion of vortex lines have not
  explicitly been considered previously (except for brief comments in
  \cite{berry2000,mrdthesis}), as the form (\ref{eq:torsion}) of
  $\tau$ is considerably more complicated than curvature $\kappa$: it
  depends on cubic combinations of fields and their derivatives
  (rather than quadratic combinations for curvature), which poses
  challenge for analytic evaluation by Gaussian techniques.
  Furthermore, the expression depends on first, second and third
  derivatives of field quantities, so the simplifications by symmetry
  and choice of coordinate system used in~\cite{berry2000} cannot
  easily be applied, because third order derivatives are correlated
  with first order derivatives.

  We instead consider the torsion PDF via Monte Carlo integration on
  the ensemble of the Gaussian model,
  \begin{equation}
    P(\tau) = \langle \tau(\vec{V}) \delta(\tau(\vec{V}) - \tau)   P(\vec{V}) \rangle_{\vec{V}}~,
  \end{equation}
  where the average is the `dislocation average' of~\cite{berry2000}
  and the vector of Gaussian random field quantities $\vec{V}$ consists of all relevant
  field derivatives from (\ref{eq:torsion}).  In the model these are
  each Gaussian distributed, and the probability $P(\vec{V})$ includes
  all of their correlations and correctly-weighted variances.  This
  PDF is shown in Figure \ref{fig:fs_pdfs}.

  The shape of the resulting numerical torsion PDF is distinctly
  different to that of the the curvature as it is a signed quantity,
  symmetric about $\tau=0$ (which is also therefore the peak of the
  distribution, and the ensemble is symmetric under $\tau \to -\tau$).
  The first moment of the unsigned distribution is
  \begin{eqnarray}
    \langle |\tau| \rangle = (6.19\pm0.09) / \lambda~.
  \end{eqnarray}
  The torsion PDF decays much more slowly than that of the curvature,
  reflecting its quantitative instability when $\kappa \approx 0$.
  The high-$\tau$ scaling from the Monte Carlo integration is shown in
  the inset to Figure \ref{fig:fs_pdfs} (a), with linear fit $\log
  P(\tau)=(-3.01\pm0.01)\tau$, suggesting that as $\tau \to \infty,$
  the probability density scales as $\tau^{-3}$.  We note that this
  behaviour occurs only when $\tau \gg \lambda;$ the fit becomes
  stable only for $\tau\gtrsim 100 / \lambda$.  If, by analogy with
  $\kappa_{\mathrm{c}},$ there is some characteristic torsion
  $\tau_{\mathrm{c}}$ determining the characteristic scaling of the
  distribution, then $\tau_{\mathrm{c}} \gg \kappa_{\mathrm{c}}.$

  The $\tau^{-3}$ scaling means that the second and higher moments of
  $\tau$ diverge, as suggested in~\cite{mrdthesis}.  It is hard to
  identify by eye what the analytic form of $P(\tau)$ might be; a
  range of PDFs with different analytic forms (including Cauchy-like
  distributions) occur as the distribution of various measures of
  vortex core twist and twirl \cite{dennis2004}), but we have not been
  able to directly fit any of these families with \emph{ad hoc}
  parameter choices to the Monte Carlo torsion distribution.

  The torsion PDF recovered from our numerical experiments is shown in
  Figure \ref{fig:fs_pdfs} (for $|\tau|$).  The fit to
  the Monte Carlo result is again reasonable.  The distribution from
  the simulations is shifted towards $\tau=0,$ similarly to the
  curvature; this shift is more pronounced here so the distribution is
  more significantly distorted from the Monte Carlo result.  This
  reflects the numerical instability in measuring the torsion in our
  simulations; since the torsion may vary significantly even when the
  line is almost straight, it is highly sensitive to both numerical
  noise in the recovered curve and to our interpolation methods to
  remove this distortion (Section \ref{sec:numerics}).  In practice it
  is not feasible to recover the torsion PDF more accurately, though
  our result is stable to alternative choices of interpolation parameter.

  The first moment of the recovered distribution is
  \begin{eqnarray}
    \langle |\tau| \rangle = (6.00\pm0.06) / \lambda~.
    \label{eq:torsqcorr}
  \end{eqnarray}
  Unsurprisingly, the visual shift to low torsions translates to a
  lower moment, and the match to the Monte Carlo integral is less good
  than that of the curvature to the analytic result.

  The Monte Carlo integration also makes accessible the joint
  curvature-torsion PDF, shown in Figure \ref{fig:fs_pdfs}(b).  It is
  consistent with the apparent distributions of Figure
  \ref{fig:fs_boxes} and with the anticipated anticorrelation of
  $\kappa$ and $\tau$.  The curvature PDF is almost recovered on the
  symmetry line $\tau = 0,$ with only a small shift in favour of
  higher curvatures.

  As with curvature, the class of small vortex loops on the wavelength
  scale does not have a typical torsion distribution: as they have
  higher than average curvature, their torsion is correspondingly
  smaller, consistent with their being approximately planar as
  anticipated in \cite{berry2007}.

  \subsection{Correlations of direction cosine, curvature and torsion}
  \label{correlation}

  Although the probability density functions for curvature and torsion
  are consistent with what is seen by eye in Figure
  \ref{fig:fs_boxes}, they do not describe how curvature and torsion
  vary {\em along} the vortex line; we consider here some simple
  statistical lengthscales along the length of a vortex curve.

  As we will discuss in more detail in Section \ref{sec:scaling}, at
  long lengthscales vortex lines in random wave tangles approximate
  random walks.  As is common in studies of Brownian random walks
  emerging from physical processes, we use the \emph{persistence
    length} $L_{\rm{p}}$ is a natural scale for classifying long-scale
  behaviour \cite{orlandini2007}.  It is defined from the
  approximately exponential decay of the {\em direction cosine
    correlation} in random walks (i.e.~correlation of tangent vectors
  $\vec{T}(s)$ with respect to arclength $s$), viz.~
  \begin{equation}
    \langle \vec{T}(s)\cdot\vec{T}(s+\Delta s) \rangle \appropto \exp(-\Delta s / L_{\rm{p}})~,
    \label{eq:persistence}
  \end{equation}
  where $\vec{T}(s')$ is the tangent to the curve at the point
  labelled by $s',$ parametrised by arclength.  As we will describe,
  the relevant $\Delta s$ here is typically longer than any relevant
  lengths that can be extracted from the curve analytically by Taylor
  expansion (unlike curvature and torsion, which are derived at a
  single point via high derivatives).

\begin{figure}
    \centering
    \includegraphics{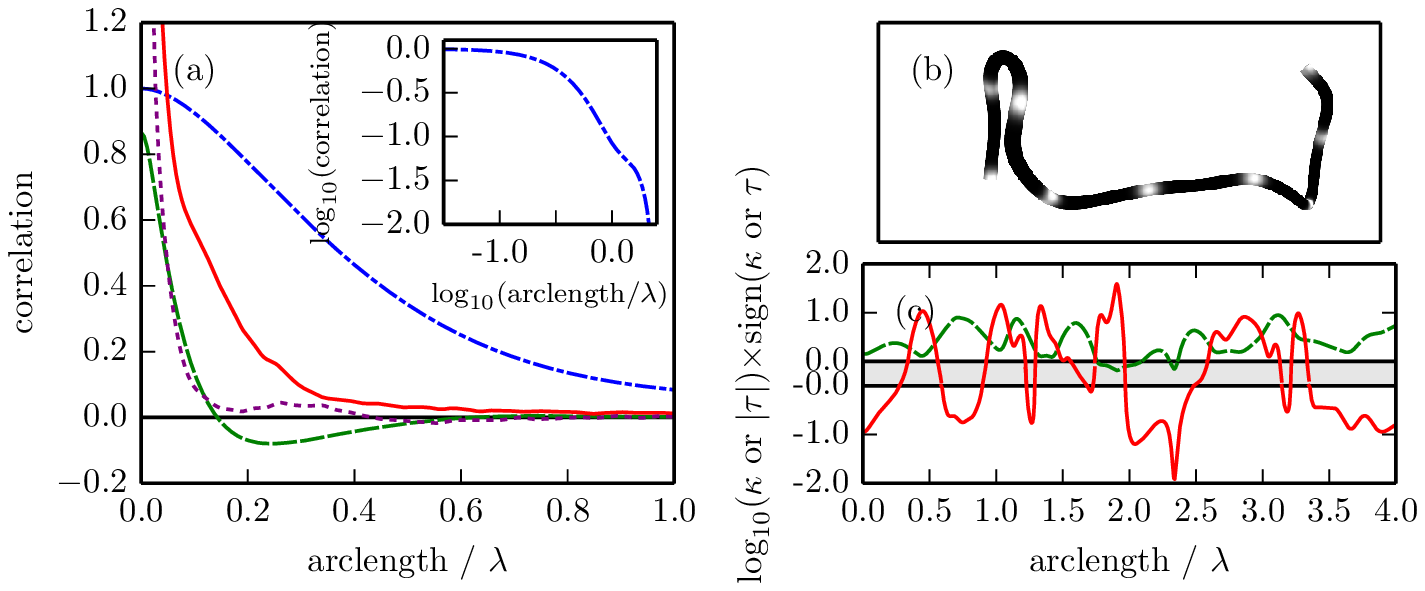}
  \caption{
    Variation of direction cosine, curvature and torsion with respect to arclength $s$ along random vortex lines.  
    (a) Correlation functions: direction cosine (\ref{eq:persistence}) (\textcolor{paperblue}{\rule[0.4ex]{1.9mm}{0.5mm}\rule[0.4ex]{0.3mm}{0pt}\rule[0.4ex]{0.6mm}{0.5mm}}), curvature correlation$-1$ (\ref{eq:curvaturecorrelation}) (\textcolor{paperdarkgreen}{\rule[0.4ex]{1.25mm}{.5mm}\rule[0.4ex]{0.3mm}{0pt}\rule[0.4ex]{1.25mm}{.5mm}}), signed torsion    correlation (\ref{eq:torsioncorrelation}) (\textcolor{paperred}{\rule[0.4ex]{3mm}{.5mm}}) and unsigned torsion correlation$-1$ (\textcolor{paperdarkpurple}{\rule[0.4ex]{0.5mm}{.5mm}\rule[0.5ex]{0.3mm}{0pt}\rule[0.4ex]{0.5mm}{.5mm}\rule[0.5ex]{0.3mm}{0pt}\rule[0.4ex]{0.5mm}{.5mm}\rule[0.5ex]{0.3mm}{0pt}\rule[0.4ex]{0.5mm}{.5mm}\rule[0.5ex]{0.3mm}{0pt}}).
    (b) shows a typical short vortex segment taken from our numerical experiments, coloured with a white dot at every $0.5\lambda$ arclength distance.
    (c) shows the varying curvature and torsion along the segment of (b). 
    The scale of (c) is the signed logarithm $\log_{10}(\kappa$~or~$|\tau|)\times\rm{sign}(\kappa$~or~$\tau)$, except in the shaded area which tracks the raw $\kappa$ or $\tau$ between $\pm1/\lambda$.
    \label{fig:fs_correlations}
  }
\end{figure}

  This direction cosine correlation is plotted in Figure
  \ref{fig:fs_correlations}.  As $\Delta s \to 0$ the correlation
  function is determined by the mean square curvature through
  application of the Frenet-Serret relations (as in
  \ref{appendix:fs}),
  \begin{equation}
    \langle \vec{T}(s)\cdot\vec{T}(s+\Delta s) \rangle = 1 - \frac{1}{2}\langle \kappa^2 \rangle(\Delta s)^2 + O(\Delta s)^4~.
  \end{equation}
  From Figure \ref{fig:fs_correlations}, this gives $\langle \kappa
  \rangle = (3.2\pm0.1)/\lambda$, consistent with the result of
  (\ref{eq:curvaturemoment}).

  When $\Delta s$ is larger, however, the correlation does \emph{not}
  apparently have the simple exponential shape of
  (\ref{eq:persistence}), decaying instead less rapidly (Figure
  \ref{fig:fs_correlations}(a) inset).  This is because the vortex
  space curve is still smooth (and not random) on the scale of
  wavelength: as $\Delta s$ increases, the randomly varying curvature
  and torsion the curve's direction, but they do so with a more
  complicated distribution.  The width of the function still gives an
  estimate for the correlation length scale; its second moment is
  $(0.590\pm0.002)\lambda$, and tangents become totally decorrelated
  over roughly the wavelength scale.

The varying of direction cosine with arclength along a vortex line can
be compared directly that of the curvature and torsion, i.e.~the
correlation functions~
\begin{eqnarray}
  C_\kappa(\Delta s) &\equiv \frac{\langle \kappa(s) \kappa(s + \Delta s)\rangle}{\langle \kappa(s) \rangle^2}~,
  \label{eq:curvaturecorrelation}\\
  C_\tau(\Delta s) &\equiv \frac{\langle \tau(s) \tau(s + \Delta s) \rangle}{\langle |\tau(s)| \rangle^2}~,
  \label{eq:torsioncorrelation}
\end{eqnarray}
These are shown in Figure \ref{fig:fs_correlations}(a).  The curvature
correlation is well-behaved, with vertical intercept matching
(\ref{eq:curvaturesqmoment}).  The torsion is noisier, and diverges as
$\Delta s\to0$.

The decay lengths of (\ref{eq:curvaturecorrelation}) and
(\ref{eq:torsioncorrelation}) give an indication of the length of
vortex segment over which curvature and torsion become decorrelated.
Both decay over a distance much shorter than a wavelength; in the case
of the curvature, the second moment of $(0.055\pm0.001)\lambda$ sets
a rough lengthscale of correlation.  Since the torsion correlation
diverges, we instead fit its shape beyond $\Delta s=0.1$ to an
exponential with decay length $(0.24\pm0.03)\lambda$.

An alternative measure of the torsion correlation distance comes from
the \emph{unsigned} torsion correlation, with $\tau$ replaced by
$|\tau|$ in (\ref{eq:torsioncorrelation}), also shown in Figure
\ref{fig:fs_correlations}(a).  As with the torsion, it diverges as
$\Delta(s)\to0$, but beyond $\Delta(s)\approxeq0.04\lambda$ it fits
well to an exponential with decay length $(0.08\pm0.01)\lambda$.  This
sets a correlation lengthscale for torsion similar to that of the
curvature.

These correlation curves and their associated lengthscales describe
different features of the random filament, as in Figure
\ref{fig:fs_boxes}.  The persistence length quantifies the arclength
interval of approximately fixed tangent direction along the curve as
it undulates; this is is the length along which a typical vortex line
is (almost) straight.  In between these sections are shorter
intervals, whose length is the curvature correlation length, at which
this direction changes rapidly.  There are other even shorter
intervals of high torsion (often when the curve is nearly straight)
where the plane of curvature varies rapidly.

Figure \ref{fig:fs_correlations}(b-c) show a segment of a vortex curve
randomly sampled from our numerical experiments, along with the
curvature and torsion along its length, demonstrating how the
correlation functions are expressed. Both exhibit small scale peaks,
with widths following the correlation lengths above.  These peaks are
often strongly related to regions of hyperbolic interchange where two
vortices approach closely, the hyperbolic analogues of the elliptic
small loops previously mentioned to occur at small scales \cite{nye2004,berry2007,dennis2008_2}.
As with the loops, the smoothness of the field demands that these
regions be highly curved but almost planar.

\section{Long-range scaling and self-similarity of vortex tangle}
\label{sec:scaling}

In the previous Section, we considered the geometry of individual
vortex lines in random waves at a scale of a wavelength or less, and
found good agreement at this local level between our numerical
experiments of random waves in periodic 3D cells and the isotropic
Gaussian random wave model.  However, such an approach cannot be
directly extended to describe the geometry of the vortex tangle at
scales beyond the wavelength -- local methods, based on Taylor
expansion, require too many terms to determine what quantities should
be calculated, and even the probability distribution of torsion could
not be calculated analytically.  Following the approach
of~\cite{oholleran2008} in studying random 3D speckle fields
satisfying the 2+1 Schr\"odinger equation (paraxial equation), we
instead investigate the statistical self-similarity of the random
vortex curves, and of the tangle: are they \emph{fractal}, what are
their statistics, and how do these compare with other models at the
large scale?

Scaling and fractality has been well investigated in other models
involving tangles of random filaments.  In \cite{vachaspati1984}, it
is shown that discrete $\mathbb{Z}_3$ phase vortices in a cubic
lattice model (as a certain simple caricature of cosmic strings in the
early universe) behave as Brownian random walks, and observe that the
assumption of a scaling continuum across longer lengthscales requires
a certain distribution of long loop lengths.  Similar results are
found for vortex tangles in random optical waves (solving the 2+1
Schr\"odinger equation)~\cite{oholleran2008}, including Brownian
fractal scaling in the radius of gyration of vortex
loops~\cite{oholleran2009}.  It is also possible to determine the
box-counting fractal dimension of the full vortex tangle, as performed
for simulations of evolving superfluid
turbulence~\cite{kivotides2001}, itself an extension of calculations
for vortices in a classically turbulent field~\cite{vassilicos1996}.
We compare all of these quantities within our own random wave model
simulations (i.e.~random solutions of the 3D Helmholtz equation whose
periodic vortex cells are cubes), verifying that the fractal scalings
of our vortices at long lengthscales are consistent with those in the
3D optical simulations~\cite{oholleran2008}.  These results are thus
the 3D counterparts to the numerous random walk and percolation
properties discussed for nodal lines in real 2D random waves as a
model for quantum chaos~\cite{bogomolny2002}.

Since the total vortex line length per simulated cell is finite,
individual filaments are always closed, but may wrap around the
periodic boundaries one or more times before closing.  If the periodic
cells are imagined as tiling infinite 3D space, such lines, on
unwrapping, are infinite and periodic; when considered in the 3-torus,
they have nontrivial homology.  Other lines which do not wrap are
simple loops, although they may span many cells.  Taking a single
period as the total arclength of a line, the total length of
individual vortex lines varies from less than a wavelength in the
smallest loops to the longest lines whose length is of comparable to
the total arclength in the cell.  These longest lines span many cells,
having lengths of over $40 000\lambda$ in periodic volumes with side
length just $21.7\lambda$.

The self-similarity of a single vortex curve can be determined by
comparing, for one point on the curve fixed and the other varying
along the curve, the arclength $s$ between the pairs of points to the
(unwrapped) Pythagorean straight line distance $R$
\cite{oholleran2008, vachaspati1984},
\begin{equation}
  \langle R \rangle \sim C s^n~,
  \label{eq:linescaling}
\end{equation}
where $C$ is a constant and $n^{-1}$ is the fractal dimension of the
line: $n = 1$ for a smooth line, $n = 1/2$ for a Brownian random walk,
and $n = 0.588$ for a self-avoiding random walk~\cite{guillou1977}.
The range at which (\ref{eq:linescaling}) applies should be larger
than the persistence length of the curve (it is smooth at smaller
distances), and less than the overall length $L$ of the curve (beyond
which it repeats, either as a loop or a periodic line).

\begin{figure}
  \begin{center}
    \includegraphics{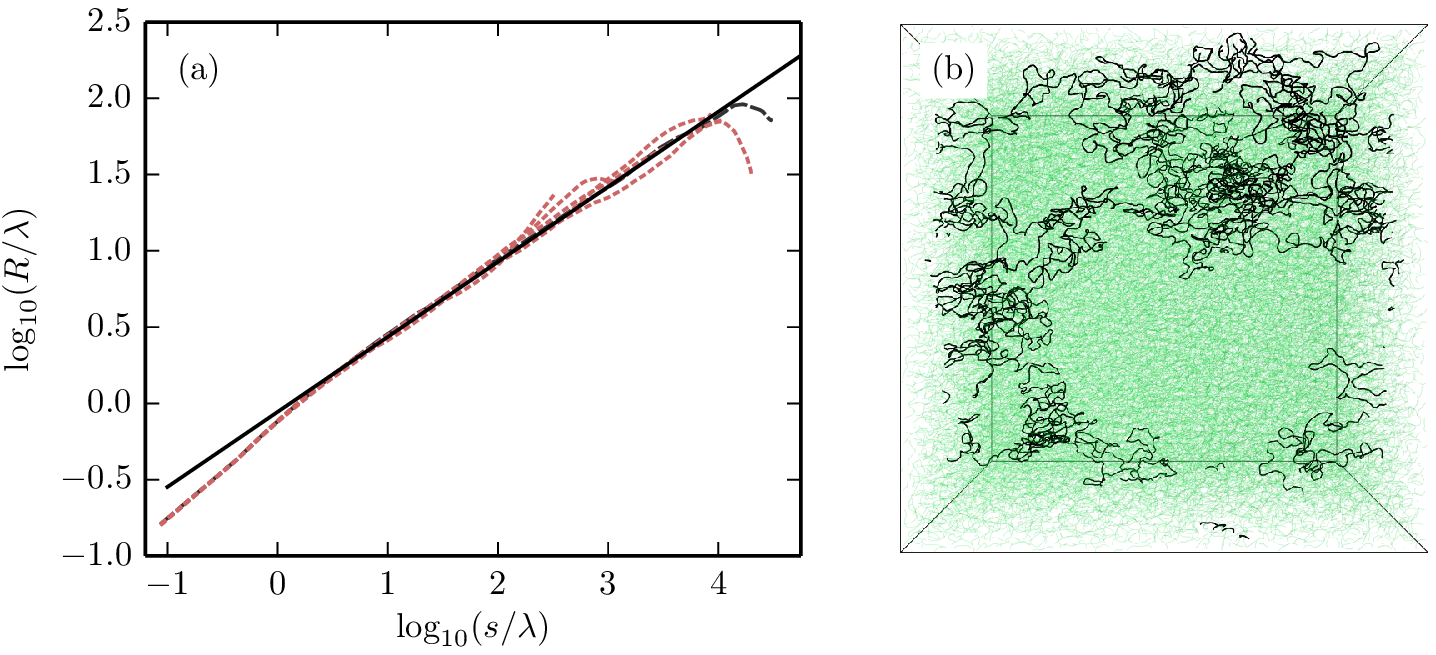}
  \end{center}
    \caption{ 
      The scaling of straight line distance between vortex points against the arclength distance between them. 
      (a) the log of the straight line distance $R$, averaged over 1500 pairs of points on the vortex curve separated by arclength $s$, against $\log_{10}(s)$. 
      Between $0.5 \le \log_{10}(s) \le 1.3$ the curves are fit to the same straight line (\rule[0.4ex]{3mm}{.5mm}) arising as the average gradient from $3000$ individual curves, shown as the continuous black line with gradient $0.504\pm0.002$.
      Also shown are 5 examples of these curves (\rule[0.4ex]{0.7mm}{.25mm}\rule[0.25ex]{0.3mm}{0pt}\rule[0.4ex]{0.7mm}{.25mm}\rule[0.4ex]{0.3mm}{0pt}\rule[0.4ex]{0.7mm}{.25mm}), alongside a single much longer curve (\rule[0.4ex]{1.25mm}{.25mm}\rule[0.4ex]{0.3mm}{0pt}\rule[0.4ex]{1.25mm}{.25mm}) almost three orders of magnitude longer than the fitted region but with the same fractal scaling persisting at larger scales. 
      In (b), a sample periodic line of length $401\lambda$ (in the middle of the length distribution), wrapped within its periodic cell, with several other lines and hundreds of loops (green).
      \label{fig:pythagorean}}
  \end{figure}

  Figure \ref{fig:pythagorean} shows this scaling for our simulated
  vortex filaments from the larger cells averaging the Pythagorean
  distance $R$ over 1500 pairs of points on each curve for each
  arclength separation $s$, giving a characteristic scaling
  relationship for each curve.  

  To limit contamination from (periodic) boundary effects, we consider
  a range of $R$ constrained by the
  cell size $0.5 \le s \lesssim 21.7 \lambda,$
  and averaging over $3000$ curves with a range of overall lengths
  (such as the sample in Figure \ref{fig:pythagorean}(b)), we find $n
  = 0.504\pm0.002$.  This is indicative of Brownian fractality,
  similar to the results of~\cite{oholleran2008}.  The lower limit for
  the Brownian scaling is around $s \approx 3.5\lambda$ (somewhat
  higher than the persistence length $L_{\rm{p}} = 0.590\lambda$
  described above), indicating that even when $s > L_{\rm{p}},$ the
  smoothness of the field affects the line's statistical
  self-similarity. The scaling at lower arclengths is $n \approx
  0.63\pm0.02$, although it is not clear what determines this exponent;
  our sampling resolution does not permit a resolution lower than
  about $0.1\lambda.$

  We also fit $10$ of the longest lines over their full length, not
  just within the span of a single cell; these vortex curves have
  lengths of at least $10^4\lambda$.  The exponent for these lines is $n =
  0.493\pm0.03$ over three decades (and over a range significantly
  longer then the side length of the cell), with a single example
  shown in Figure~\ref{fig:pythagorean}(a).  This scaling behaviour
  continues the initial trend and appears highly typical, limited only
  by the size of the periodic cell.  The characteristic hooked shape
  at the longest scales is a consequence of the periodic boundary
  conditions and the fact that cells tend to contain only one such
  very long line.  This may be understood heuristically as follows.
  Each periodic cell must have the same vortex flux entering and
  leaving opposite faces.  This constrains the paths of infinite
  periodic lines; for every `loop' through the periodic boundaries,
  the same line or a different one must make an equal but opposite
  loop for the total flux to sum to $0$.  A long, unwrapped periodic
  line cannot typically roam far from any reference point, as the
  other lines in the cell must counter its flux.  Since the other
  lines are typically much shorter, the long line cannot venture far
  and instead behaves almost like a loop, typically returning to a
  cell near its starting point.  This manifests as the observed hook
  shape, in a similar manner to the way a loop losing its fractality
  at arclengths approaching its overall length by being forced to
  close within a finite distance.

\begin{figure}
    \begin{center}
      \includegraphics{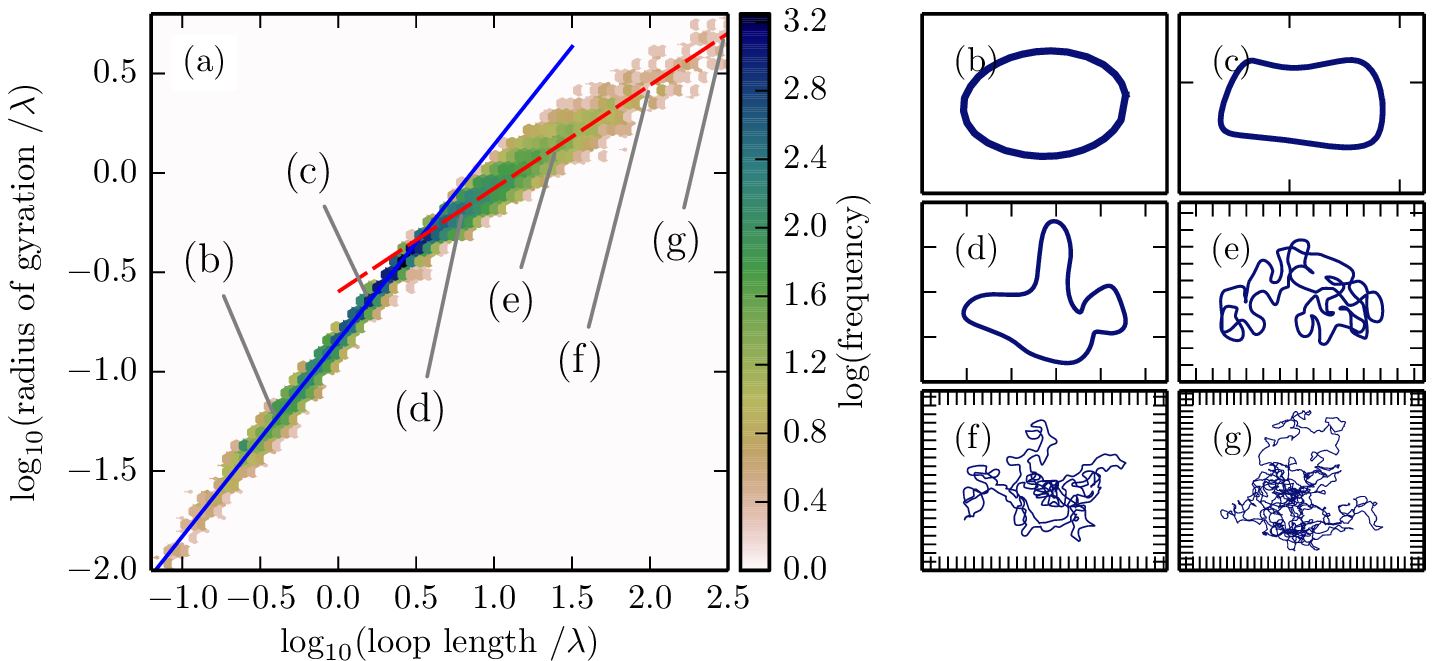}
    \end{center}
    \caption{Scaling of random vortex loops.  (a) Log-log plot of
      radius of gyration $r_{\rm{g}}$ against loop length $L,$ for
      39487 loops from 650 simulated cells of side length
      $7.8\lambda$.  The fit for the lower values
      (\rule[0.4ex]{3mm}{0.5mm}) has gradient $0.98\pm0.01$, while the
      fit for higher values
      (\rule[0.4ex]{1.7mm}{0.5mm}\rule[0.4ex]{0.3mm}{0mm}\rule[0.4ex]{1.7mm}{0.5mm})
      has gradient $0.52\pm0.01$, with the cutoff between regimes at
      $(3.5\pm0.1)\lambda$.  (b-c) show examples of loops randomly
      selected at different length scales.
      \label{fig:rog} 
    }
  \end{figure}

  The fractal self-similarity of closed loops may also be determined
  by comparing the loop's overall length $L$ against its radius of
  gyration $r_{\mathrm{g}}$ (i.e.~the root mean square distance of all
  points on the loop from its centroid).  At the smallest scale, an
  analysis based on local Taylor expansion shows that small vortex
  loops typically resemble small puckered ellipses, whose radius of
  gyration scales linearly with length, i.e.~$r_{\mathrm{g}} \propto
  L$~\cite{berry2007}.  However, for longer loops to have the
  characteristics of Brownian walks we expect that $r_{\mathrm{g}}
  \propto L^{1/2}.$ A log-log plot of loop length against radius of
  gyration for simulated vortex loops is shown in Figure
  \ref{fig:rog}, clearly demonstrating these two regimes, again with a
  rather sharp transition at $L \approx 3.5\lambda.$ These loops are
  are taken from cells of side length $7.8\lambda$.  The smaller size
  is numerically optimal for tracing vortices in many cells in
  parallel, necessary since the radius of gyration cannot be averaged
  across a single loop, so many data points are necessary to check the
  scaling.

  The fit beyond $L\approx 3.5\lambda$ has gradient $0.52\pm0.01$,
  consistent with the $0.5$ expected for random walks, though there is
  wide variation for individual loops.  Below the transition the fit
  is instead to a gradient of $0.98\pm0.01$, consistent with the
  scaling of simple ellipses.  Figure \ref{fig:rog}(b-g) shows
  examples of randomly selected loops across all these scales; small
  loops such as (b) demonstrate the small scale ellipse shape required
  by the smoothness of the field, while approaching the gradient
  cutoff introduces higher order terms that disrupt this shape as in
  (c).  Beyond the cutoff the loops become tangled, no longer limited
  by the local smoothness of the field.

\begin{figure}
    \begin{center}
      \includegraphics{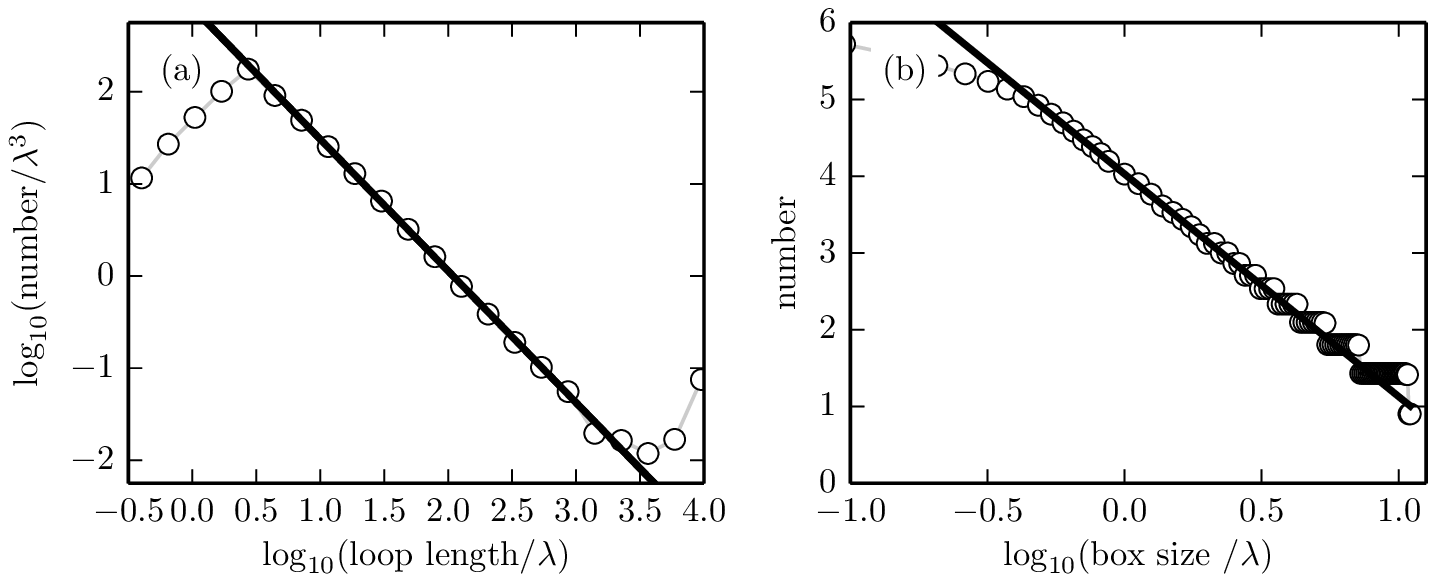}
    \end{center}
    \caption{ 
    \label{fig:boxcounting}
      Scaling quantities for vortex curves. 
      (a) Log-log plot of loop number (of fixed length) against the loop length, with a fit to gradient $-1.447\pm0.003$, from $1245899$ loops in $4616$ simulation cells with $N=9$. 
      (b) Deducing the box counting dimension of (\ref{eq:boxcounting}), with fit to the gradient $2.90\pm0.01$, averaging over $10$ simulation cells with $N=25$.
      \label{fig:52scaling}
    }
  \end{figure}

  As discussed in~\cite{vachaspati1984}, a truly scale invariant
  distribution of curves must appear the statistically the same at any
  given lengthscale.  For Brownian curves, this suggests the number
  density of curves $N(L)$ of length $L$ should scale with $L$
  according to~\cite{vachaspati1984,oholleran2009}
  \begin{equation}
    \label{eq:52scaling}
    N \propto L^{-3/2}.
  \end{equation}
  Figure \ref{fig:52scaling} shows the corresponding scaling from our
  numerical experiments.  Beyond a loop length of $3.5\lambda$,
  $N(L)$ decays as $L^{-(1.447\pm0.003)}$.  This is not fully
  consistent with our expectation, which we believe to be mainly an
  artefact of periodicity; as with the discussion of the numerical
  vortex density in Section \ref{sec:introduction}, a difference
  between our periodic system and the ideal isotropic case is a
  requirement that lines close on relatively short scales, rendering
  the ideal continuum distribution of loops inaccessible.  This same
  effect is responsible for the peak at maximal $L$; in the continuum
  model, arbitrarily long loops would occur, but instead the
  periodicity constraint limits the total loop length to approximately
  the total arclength in a cell.

  The preceding scaling measures apply to individual vortex curves.
  It is also possible to examine the self-similarity of the tangling
  \emph{neighbourhood} of any point in the vortex field by means of
  the box counting dimension of the tangle~\cite{falconer1997}.  This
  is obtained by tiling the cell with \emph{boxes} of side length
  $\delta$, and considering the number of boxes $N_{\rm{b}}(\delta)$
  containing a vortex line segment as a function of $\delta$.  As
  above, we consider this in the range of large $\delta,$ as the
  tangle is smooth in the range $\delta \approx \lambda.$

Following~\cite{vassilicos1996}, the box counting fractal dimension
$n_{\rm{b}}$ is defined as the scaling exponent in the expression
\begin{equation}
  \label{eq:boxcounting}
  N_{\rm{b}}(\delta) \propto \delta^{n_{\rm{b}}}~
\end{equation}
over some range of $\delta$ (which we expect to be larger than
$\lambda$); in the limit $\delta\to0$, $N_{\rm{b}}$ approaches the
number of vortex sampling points; when $\delta \lesssim \lambda,$
$n_{\rm{b}} \approx 1,$ since the vortex tangle consists of lines.
However, when $\delta$ is very large (say the side length of the
periodic vortex cell), $N_{\rm{b}}$ must simply equal 3 as every box
will include at least one vortex segment.  Any fractal region, if it
exists, occurs between these two extremes.

$n_{\rm{b}}$ is therefore a measure of the proportion of space (at a
scale given by $\delta$) which is filled by the curves.  The box
counting scaling for our numerical tangles is shown in Figure
\ref{fig:boxcounting}; the fit is to $n_{\rm{b}}=2.90\pm0.01$ across a
full decade of $\delta$: clearly $n_{\rm{b}} = 3$ indicates space is
homogeneously filled at these scales.  This space-filling regime in
fact begins at around $\delta\approx0.6\lambda$, well below the
fractality scale of the other measurements and roughly on the scale of
the persistence length.  Thus the range in which the box-counting
dimension is unity appears negligible.

In fact, this lower limit of $0.6 \lambda$ is close to the reciprocal
square root density of points where vortex lines cross an arbitrary
plane, of $\sqrt{3/2\pi} \lambda \approx 0.69 \lambda;$ this is a
measure of the mean spacing of vortex points.  The fact that the
space-filling regime begins at this scale seems to be an indicator of
the previously discussed rigidity and regularity of random fields
satisfying the Helmholtz equation \cite{foltin2004,bogomolny2007}.
This property has been related in 2D to the infinite
screening length of the vortex points as topological charges
\cite{kessler2008,berry2000}.  Given that the screening length of
distributions of wavenumbers (such as a Gaussian), we anticipate that
vortices in these fields would display $n_{\rm{b}} = 3$ only at larger
values of $\delta$.

Box counting fractality has also been investigated numerically in the
vortex tangle of superfluid turbulence, where the scaling is directly
affected by the dynamics of the flow and normal fluid
interaction~\cite{kivotides2001}.  The fractal dimension depends on
the vortex line density, but ranges only between $1.4$ and $1.7$ over
a wide range of physical parameters.  This is a strong contrast to
vortices in random waves; although the vortex bulk is fractal, the
lower dimension implies the line density is not distributed locally
isotropically, instead surrounding vortex voids at all length scales.
This is not evident in the self-similarity scaling of a single vortex,
but is revealed when considering the scaling of the tangle as a whole.

  \section{Discussion}
  \label{sec:discussion}

  We have described various features of the random wave vortices/nodal
  lines in (almost)-isotropic superpositions of random waves
  satisfying the 3D Helmholtz equation, extracted mainly from
  numerical experiments in cubic 3D cells with periodic boundary
  conditions.  Along a single vortex curve at scales of around a
  wavelength, we have seen that the probability distributions of
  curvature and torsion extracted from the simulations agrees well
  with the results of the isotropic Gaussian random wave model, and
  have identified numerically the different lengths of order $\lambda$
  at which various scalings occur.  At larger length scales, we have
  verified that the tangle homogeneously fills space, but individual
  lines look like Brownian random walks and the distribution of closed
  loop lengths appears scale-invariant.

  Many questions naturally arise from this numerical study. Most
  importantly, given the simplicity of this random wave model, might
  it be possible to derive any of these properties rigorously from the
  Helmholtz equation?

  The geometric properties of vortex lines arise purely from
  interference, and the probability distributions of curvature and
  torsion are not difficult to extract (at least numerically) using
  the random wave model.  Vortex lines in isotropic random functions
  with different power spectra were found to have similar properties
  to random Helmholtz waves in \cite{berry2000} (such as the same
  curvature distribution (\ref{eq:curvature_pdf}) but with different
  $\kappa_{\mathrm{c}}$); it is not clear whether similar PDFs lead to
  similar conformations of vortex lines, or whether they depend more
  subtly on correlations along vortex lines.  According to the
  fundamental theorem of space curves, a curve's shape is determined
  uniquely by the curvature and torsion as functions of arclength.
  One might expect that the random walk behaviour arises from the
  vanishing of the various correlation functions along the vortex
  lines, but do these have additional subtle features which
  characterise them as random wave vortex curves? For instance,
  preliminary study of a different class of closed curves, specified
  by random finite Fourier series in Cartesian directions, look rather
  different to random vortex curves on visual inspection, yet have
  qualitatively similar curvature and torsion PDFs, and direction
  cosine, curvature and torsion correlation functions.  Therefore,
  what geometric properties of random vortex lines characterise their
  geometry in particular?

  Questions of the large scale \emph{topology} of random vortex lines
  in the simulations are strongly affected by the periodic boundary
  conditions, in that most vortex length is in lines of nontrivial
  homology which wrap around the cell a nonzero number of times (which
  of course cannot occur in infinite cells).  Such nontrivial homology
  lines appear to consume the majority of arclength in a given cell;
  in our cells with $N=9$, they make up on average $(83.1\pm0.2)\%$ of
  the total arclength in a cell, though this fraction varies widely
  between individual cells (it is not uncommon in fact for a cell only
  to contain lines of zero homology).
  This fraction is commensurate with $73\%$ of lines in the optical
  model~\cite{oholleran2009}, or $81$-$86\%$ in models of cosmic
  strings~\cite{vachaspati1984}.

  In fact, we find that each vortex cell typically contains one very
  long line of nontrivial homology, somewhat analogous to the
  existence of a percolating cluster in regular percolation theory
  \cite{stauffer1994}, and a small number of shorter ones.
  \cite{nahum2013}~suggests a Poisson-Dirichlet distribution for the
  distribution of line lengths is generic to a broad class of systems,
  with no distinction between lines with different homology;
  preliminary investigation suggests that our system meets this
  prediction, but we do not investigate this here.  As discussed in
  the main text, the total period length of the percolating line tends
  to be rather larger than the norm of its homology vector (i.e.~the
  Pythagorean period length).

  This local and long-range description characterises the tangled
  nodal forest in 3D complex, fixed-frequency noise.  As such, the
  quantities we have described can be used as signatures for wave
  chaos in 3D complex billiard eigenfunctions or resonances, despite
  the fact that the PDFs, correlations and scaling laws do not appear
  to be particularly idiosyncratic to random wave vortex filaments
  rather than other patterns of tangles that occur in physics.  As
  described in \cite{berry2007,dennis2008_2},
  features such as small loops and hyperbolic avoidances are
  determined by higher-order defects in the gradient of the complex
  scalar field (such as nodal lines of the complex scalar $\nabla \psi
  \cdot \nabla \psi$), suggesting important features of the vortex
  tangle are in fact described by a hierarchy of tangled filaments
  describing the complex 3D morphology of the random field.
  
  \ack

  We are grateful to S Whittington, M V Berry, J H Hannay and G P
  Alexander for discussions. This work was partially supported by NSF
  Grant No.~PHY11-25915 under the 2012 KITP miniprogram ``Knotted
  Fields.'' AJT acknowledges financial support from the EPSRC.

\providecommand{\newblock}{}

  \appendix

  \section{Frenet-Serret Geometry}
  \label{appendix:fs}

  In our geometrical analysis, we consider vortices as one-dimensional
  space curves in three dimensions.  Here we briefly review the
  details of the so-called Frenet-Serret formalism which describes the
  curvature, torsion and related quantities for a space curve in terms
  of its arclength parameter $s,$ and how they relate to our physical
  model of wave vortex filaments.  The following geometrical results
  about space curves are standard material, and can be found in
  textbooks such as \cite{moderndifferentialgeometry}.

We begin by assuming the curve is given by a vector function $\boldsymbol{\gamma}(t),$ where $t$ is some arbitrary parameter which varies along the curve.
Denoting derivatives with respect to $t$ by $\dot{\bullet},$ the $t$-derivative $\dot{\boldsymbol{\gamma}}$ points in the direction of the (unit) tangent vector $\vec{T}$ times a nonnegative quantity called the \emph{velocity} $v,$ i.e.~$\dot{\boldsymbol{\gamma}} = v \vec{T}.$
In the case where the parametrization is the arclength, $v = 1,$ and $\frac{\rmd}{\rmd s}\boldsymbol{\gamma} = \vec{T}.$

The Frenet-Serret formalism follows simply from this fact.
The derivative of the tangent with respect to the arclength, 
\begin{equation}
  \frac{\rmd\vec{T}}{\rmd s} = \kappa \vec{N},
\end{equation}
defines the unit \emph{normal} vector $\vec{N}$ (which satisfies $\vec{N}\cdot\vec{T} = 0$) and the nonnegative definite \emph{curvature} $\kappa,$ which determines the magnitude by which the tangent changes with arclength. 
A third vector, the \emph{binormal} $\vec{B}\equiv\vec{T}\times\vec{N}$, completes an orthonormal basis at every regular point on the curve (i.e.~where $\kappa >0;$ this is true almost everywhere on vortex lines in random waves).
Taking the derivative of $\vec{N}$ with respect to arclength,
\begin{equation}
  \frac{\rmd}{\rmd s}\vec{N} = \kappa \vec{T} + \tau \vec{B},
\end{equation}
defining the \emph{torsion} $\tau,$ which can be positive (if the curve bends into the positive binormal direction) or negative (negative binormal).
The shape of the curve is thus completely determined by the curvature and torsion as functions of arclength; this result is the \emph{fundamental theorem of space curves}.

If the curve is specified by a parameter $t$ which is not simply the arclength, the curvature and torsion can still be extracted from derivatives of $\boldsymbol{\gamma}$ with respect to $t$ using the chain rule,
\begin{eqnarray}
  \ddot{\boldsymbol{\gamma}} & = & v^2\kappa\vec{N} + \dot{v} \vec{T} \\
  \dddot{\boldsymbol{\gamma}} & = & v^3 \kappa \tau \vec{B} + (2v \dot{v} \kappa + v^2 \dot{\kappa})\vec{N} + (v \dot{v}+\ddot{v})\vec{T}.
\end{eqnarray}
Using the orthogonality properties of the vectors $\vec{T}, \vec{N}$ and $\vec{B},$ it is straightforward to show that
\begin{eqnarray}
  \kappa & = & \frac{|v \ddot{\boldsymbol{\gamma}} - \dot{v} \dot{\boldsymbol{\gamma}}|}{v^3} \label{eq:kappat}\\
  \tau & = & \frac{\dot{\boldsymbol{\gamma}} \times \ddot{\boldsymbol{\gamma}} \cdot \dddot{\boldsymbol{\gamma}} }{|v \ddot{\boldsymbol{\gamma}} - \dot{v} {\boldsymbol{\gamma} }|^2}, \label{eq:taut}
\end{eqnarray}

On a vortex curve, the vorticity vector $\boldsymbol{\omega} \equiv \frac{1}{2}~\mathrm{Im}~\nabla \psi^*\times \nabla \psi,$ which points along the vortex line, conveniently defines a parametrization of a vortex curve whose velocity is the modulus $\omega = |\boldsymbol{\omega}|,$ and $\boldsymbol{\omega} = \omega \vec{T}.$
Furthermore, derivatives with respect to this parametrization are represented by a convective-like derivative $\boldsymbol{\omega}\cdot\nabla.$
Appropriate substitutions of $\boldsymbol{\omega}$ and $\boldsymbol{\omega}\cdot\nabla$ for $\dot{\boldsymbol{\gamma}}$ and $\rmd/\rmd t$ in (\ref{eq:kappat}) and (\ref{eq:taut}) give the expressions (\ref{eq:curvature}) and (\ref{eq:torsion}) used in the main text.

\end{document}